\documentclass[conference,a4paper]{IEEEtran}

\IEEEoverridecommandlockouts
\usepackage{cite}
\usepackage{amsmath,amssymb,amsfonts}
\usepackage{algorithmic}
\usepackage{graphicx}
\usepackage{textcomp}
\usepackage{xcolor}
\usepackage{balance}
\def\BibTeX{{\rm B\kern-.05em{\sc i\kern-.025em b}\kern-.08em
    T\kern-.1667em\lower.7ex\hbox{E}\kern-.125emX}}

\usepackage{siunitx}
\usepackage[hidelinks]{hyperref}
\usepackage{mathtools}
\usepackage{nicematrix}
\usepackage{balance}
\usepackage{enumitem}
\usepackage{tikz}
\usepackage{textcomp}
\usepackage{braket}
\usepackage{bm}

\usepackage{float} 

\hypersetup{ %
    breaklinks=true, %
    citecolor=black, %
    colorlinks=true, %
    linkcolor=black, %
    urlcolor=blue
 }

\catcode`,\active

\catcode`\,12

\newcommand\copyrighttext{%
  \footnotesize \textcopyright 2026 IEEE. Personal use of this material is permitted. Permission from IEEE must be obtained for all other uses, in any current or future media, including reprinting/republishing this material for advertising or promotional purposes, creating new collective works, for resale or redistribution to servers or lists, or reuse of any copyrighted component of this work in other works.
  DOI: \href{https://doi.org/10.1109/APWCS71511.2026.11742921}{10.1109/APWCS71511.2026.11742921}
  }
\newcommand\copyrightnotice{%
\begin{tikzpicture}[remember picture,overlay]
\node[anchor=south,yshift=70pt] at (current page.south) {\fbox{\parbox{\dimexpr\textwidth-\fboxsep-\fboxrule\relax}{\copyrighttext}}};
\end{tikzpicture}%
}

\begin{document}

\title{Multipartite FSO/QKD with N-GHZ States for Scalable Secure Group Key Distribution} 

\author{
\IEEEauthorblockN{Yuma Kaminaga, Hoang D. Le, Cuong T. Nguyen, and Anh T. Pham}

\IEEEauthorblockA{%
Computer Communications Laboratory, The University of Aizu, Aizuwakamatsu 965-8580, Japan}
}

\maketitle

\copyrightnotice

\begin{abstract}
The emergence of sixth-generation (6G) wireless networks is driving the need for secure group key generation mechanisms resilient to quantum-computing attacks. Conventional quantum key distribution (QKD) protocols are designed for point-to-point communication and are therefore not directly applicable to group key establishment. To address this limitation, this paper proposes a scalable High-Altitude Platform (HAP)-assisted multipartite QKD network architecture that integrates $N$-GHZ-state protocols with a Time Division Multiple Access (TDMA) scheme over free-space optical (FSO) channels. The performance of the proposed system is evaluated using a Qiskit-based simulation framework that incorporates realistic physical-layer impairments of FSO links. Simulation results demonstrate that the proposed architecture significantly improves the group secret key rate (SKR) compared to conventional bipartite approaches, while effectively reducing overall network execution delay. In addition, the analysis reveals the impact of channel impairments and provides practical design insights, such as optimal beam-waist control. The proposed framework demonstrates the potential of HAP-assisted multipartite QKD as an effective security solution for future large-scale 6G wireless networks supporting secure group communications.
\end{abstract}
 

\section{Introduction}
\label{sec:introduction}
In recent years, the rapid evolution of sixth-generation (6G) wireless networks has created an urgent demand for secure group-key generation mechanisms capable of supporting simultaneous data exchange among massive numbers of distributed wireless users. Emerging 6G applications, such as immersive communications, autonomous systems, intelligent transportation, and large-scale machine-type communications, require highly secure and low-latency group communications across wide-area networks.

Conventional group key establishment protocols typically require a large number of communication rounds and computationally intensive exponentiation at each device, making them poorly suited for resource-constrained wireless network environments~\cite{hegland2006survey}. Moreover, since these schemes rely on classical computational hardness assumptions, they are fundamentally vulnerable to emerging quantum computing attacks. To address these limitations, quantum key distribution (QKD) has garnered significant attention as a promising, future-proof alternative for secure communication~\cite{scarani2009security}. Using quantum-mechanical laws, QKD enables legitimate parties to generate and share secret keys in a way that inherently detects any eavesdropping attempt.

Currently, most QKD implementations rely on optical fiber infrastructure. However, deploying optical fibers to individual wireless devices distributed across diverse environments is impractical. To overcome this limitation and enable flexible wide-area QKD deployment, high-altitude platform (HAP)-based systems over free-space optics (FSO) have attracted considerable research attention. Nevertheless, existing HAP-based QKD system mainly adopt prepare-and-measure protocols, such as BB84, or bipartite entanglement-based protocols, such as BBM92~\cite{karakosta2025free, chu2021feasibility, alshaer2024exploring, vu2020design, trinh2025optical, kaminaga2026qiskit}. Since these protocols are essentially designed for point-to-point key generation, extending them to group key distribution inevitably introduces sequential communication delays, similar to those seen in conventional classical approaches.
\begin{figure}[t]
  \centering
    \includegraphics[width=0.9\columnwidth]{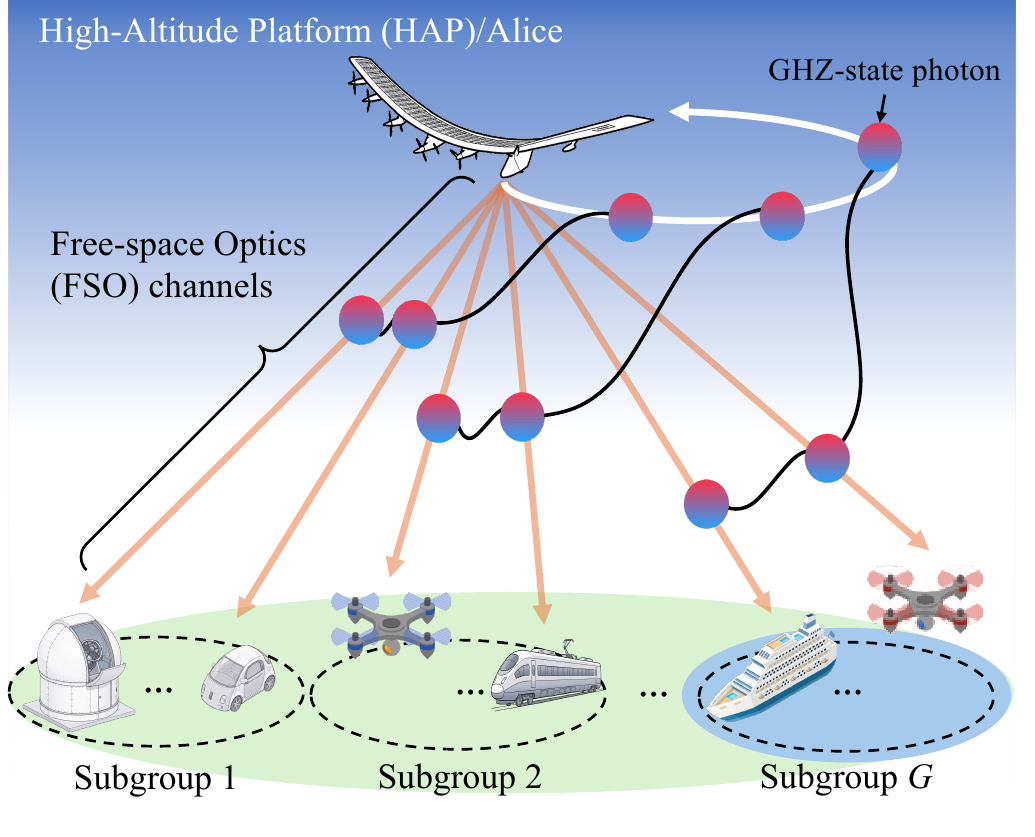}
    \caption{The considered HAP-aided FSO/QKD system for quantum group key distribution.}
    \label{fig:system_model}
\end{figure}

To overcome the inherent point-to-point limitation of conventional QKD, multipartite entanglement using Greenberger-Horne-Zeilinger (GHZ) states has emerged as a promising solution for direct group key generation \cite{epping2017multi, pickston2023conference}. Theoretically, a large-scale $N$-GHZ state could simultaneously connect all users within a single execution. However, the generation and distribution of such massive entangled states remain technologically challenging and are still largely confined to experimental demonstrations.

Motivated by this limitation, this work proposes a scalable HAP-assisted centralized NQKD network architecture that integrates practically realizable small-scale $N$-GHZ states with a time-division multiple access (TDMA) scheme. By clustering ground users into optimized small groups and sequentially distributing entangled states, the proposed framework effectively bypasses the physical limits of large-scale entanglement generation while mitigating the severe delay bottlenecks encountered in conventional pairwise QKD systems. The primary goal of this study is to develop a comprehensive Qiskit-based simulation framework that explicitly incorporates realistic FSO physical-layer impairments, including atmospheric attenuation, turbulence-induced fading, and pointing errors, together with Pauli-based quantum error models to accurately emulate quantum noise environments. Using the developed framework, we quantitatively evaluate and compare the performance of the $N$-GHZ-based protocol against conventional bipartite 2-GHZ methods under identical channel conditions. 
\section{System Description}
\label{sec:system_desp}
\subsection{System Model}
\label{sec:system_model}
\begin{figure}[t]
  \centering
    \includegraphics[width=\columnwidth]{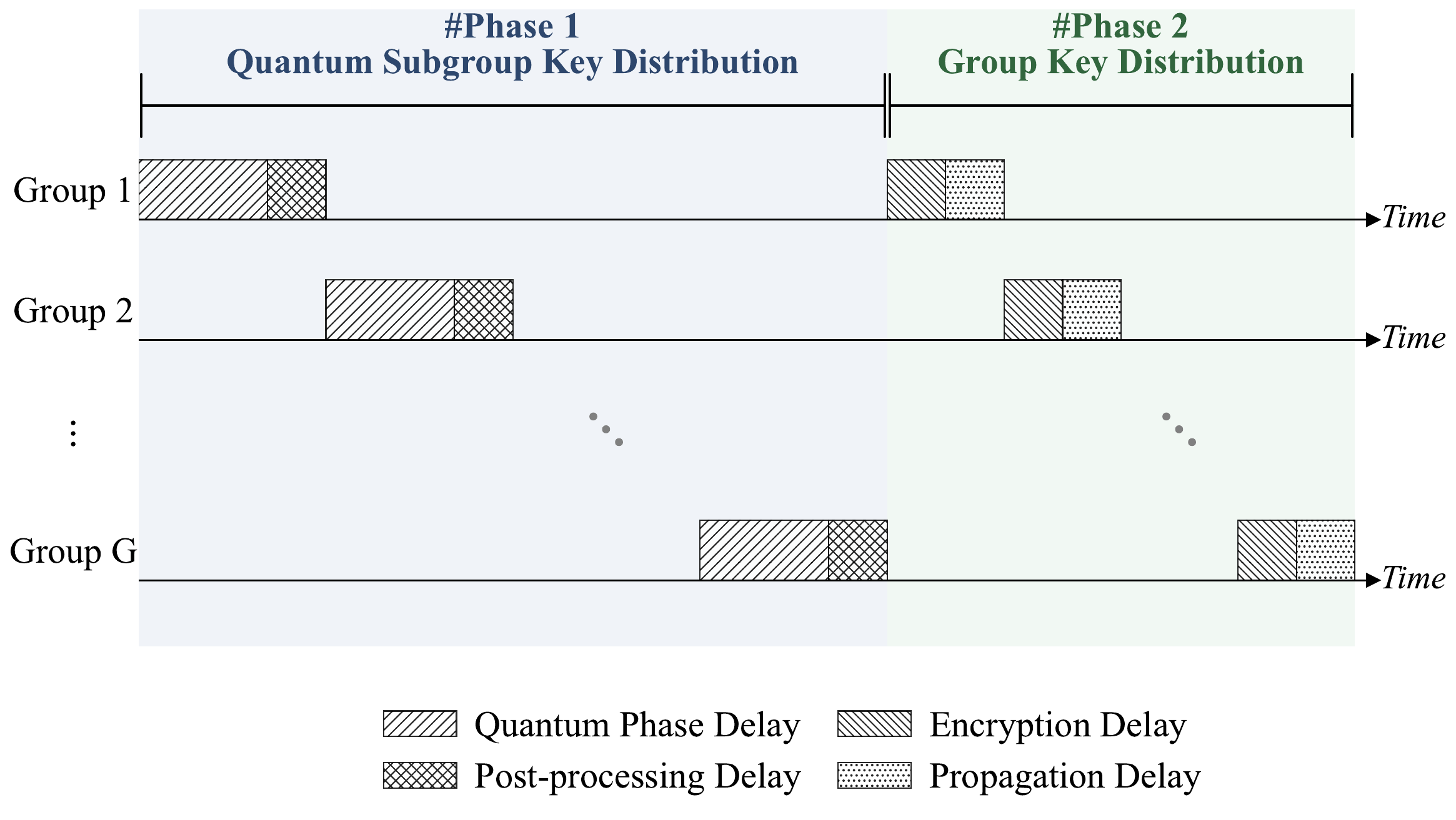}
    \caption{Timeline of two phases to establish the group secret key.}
    \label{fig:group_key_steps}
\end{figure}
Fig.~\ref{fig:system_model} illustrates the proposed HAP-assisted system, where a HAP (Alice) distributes a group key to $M$ wireless users (Bob) via FSO channels. The proposed framework consists of two phases: (1) quantum subgroup key distribution and (2) group key distribution, as depicted in Fig.~\ref{fig:group_key_steps}. In the first phase, users are divided into $M$ subgroups, and the HAP establishes a shared key with each subgroup using TDMA and the $N$-GHZ protocol. In particular, the transmission period is divided into $G$ equal time slots, each assigned to one subgroup. During each slot, Alice generates $N$-GHZ entangled photons, retains one photon, and distributes the remaining $N-1$ photons to the corresponding group users via FSO links. After a transmission duration of $\Delta t$, Alice and the users perform post-processing over a public channel to establish a subgroup key, $K_i$. 

After the first phase, HAP obtains $G$ subgroup keys, $K_i$ ($i \in \left\{ 1, \dots, G \right\}$), each shared with the corresponding subgroup. In the second phase, Alice generates a global group key $K_\text{G}$ and encrypts it for each subgroup using the XOR operation, i.e., $K_{\text{G}, i} =  K_\text{G} \oplus K_i$. HAP broadcasts $K_{\text{G}, i}$, and the users in that subgroup can decrypt the group key via $K_\text{G} = K_{\text{G}, i} \oplus K_i$. Consequently, all $M$ users and Alice share the same group key $K_\text{G}$.
The GHZ entangled state used in the protocol is expressed as follows~\cite{epping2017multi}
\begin{align}
    \label{eq:ghz_state}
    \lvert \mathrm{GHZ} \rangle = \frac{1}{\sqrt{2}} \left( \lvert 0 \rangle^{\otimes N} + \lvert 1 \rangle^{\otimes N} \right).
\end{align}

Upon receiving a photon, each user randomly selects either a rectilinear (Z) or diagonal (X) basis for measurement with probabilities $P_\text{Z}$ and $P_X = 1 - P_\text{Z}$, respectively.
The users then announce their chosen bases in a public channel and retain the sifted bits according to the GHZ state order ($N$). For $N = 2$, bits are kept only when all users choose the same basis (either $Z$ or $X$). This rule is identical to the BBM92 protocol. For $N > 2$, only bits measured entirely in the $Z$ basis by all users are retained.

\subsection{Free-space Quantum Channel Model}
\label{sec:channel_model}

We denote the transmittance of the quantum channel from HAP to the $i$-th Bob as $\eta_i$, which can be modeled as \cite{karakosta2025free}
\begin{align}
    \label{eq:free_space_efficiency}
    \eta_i = \eta_{\mathrm{atm}} \eta_{D_{\mathrm{Rx}}} \, \eta_{\mathrm{SMF}},
\end{align}
where $\eta_{\mathrm{atm}}$ is the atmospheric transmittance, $\eta_{D_{\mathrm{Rx}}}$ is the receiver collection efficiency, and $\eta_{\mathrm{SMF}}$ is the fiber-coupling efficiency. These factors are detailed as follows.

\subsubsection{Collection efficiency}
\label{sec:collection_eff}
 $\eta_{D_{\mathrm{Rx}}}$ represents the fraction of power that can be collected at the receiving telescope in the presence of turbulence-induced fading, beam wandering, and pointing error. The probability density function (PDF) of the collection efficiency can be written as \cite[(B21)]{karakosta2025free}
\begin{align}
    \label{eq:collection_efficiency_polar}
    f_{\eta_{D_{\mathrm{Rx}}}} \left( \eta_{D_{\mathrm{Rx}}} \right) &= \frac{1}{\sigma_{\mathrm{w}}^2} \int_{0}^{\infty} P(\eta_{D_{\mathrm{Rx}}}|r) \exp\left[-\frac{r^2}{2\sigma_{\mathrm{w}}^2}\right] r \, \mathrm{d}r,
\end{align}
where $r$ is the beam displacement, $\sigma_{\mathrm{w}}^2$ is the variance of the beam displacement, which combines beam wandering variance and the pointing error, or is given as \cite[(7)]{karakosta2025free}
\begin{align}
    \label{eq:tracked_wander_variance}
    \sigma_{\mathrm{w}}^2 = \Big[ \left( z\theta_{\mathrm{pe}} \right)^2 +  r_c^2 \Big] \, (1 - \eta_{tr}),
\end{align}
where 
$z$ is the propagation distance,
$\eta_{tr}$ is the tracking efficiency,
$\theta_{\mathrm{pe}}$ is the angular pointing error,
and $r_c^2$ is turbulence-induced beam wandering variance and is given as \cite[(B12)]{karakosta2025free}
\begin{align}
    \label{eqn:beam_wandering}
    r_c^2 = 7.25 \sec^3 \left( \theta_z \right) W_0^{-1/3} \displaystyle\int_{h_0}^{H} \frac{C_\text{n}^2 \left( h \right)(h - h_0)^2}{ \Xi } \mathrm{d}h,
\end{align}
where
$\theta_z$ is the zenith angle,
$\sec \left( \cdot \right)$ is the secant function,
$W_0$ is the beam waist radius,
$H$ and $h_0$ are the altitudes of HAP and the ground user relative to sea level, respectively. Moreover, $C_\text{n}^2 \left( h \right)$ is the refractive index structure parameter, and can be computed via the ground-level refractive index $C_\text{n}^2 \left( 0 \right)$ and the root-mean-square wind speed $v_\text{wind}$ as in \cite[(14)]{trinh2025optical}, and $\Xi \!\! = \!\! \left[ \left( \Theta_0 + \bar{\Theta}_0 \frac{h - h_0}{H - h_0} \right)^2 \!\!\! + \! 1.63 \sigma_\text{R}^\frac{12}{5} \Lambda_0 \left( \frac{H - h}{H - h_0} \right)^\frac{16}{5} \right]^\frac{1}{6}$, where $\Theta_0$, $\bar{\Theta}_0$, and $\Lambda_0$ can be found in \cite[(B11)]{karakosta2025free}, and $\sigma^2_\text{R}$ is the Rytov variance, which is calculated as a function of the optical wavelength $\lambda$ in \cite[(4)]{kaminaga2026qiskit}.

Moreover, $P(\eta_{D_{\mathrm{Rx}}}|r)$ is the conditional PDF of the collection efficiency given the beam displacement $r$, which can be approximated by the truncated log-normal distribution and can be written as \cite[(B19)]{karakosta2025free}
\begin{align}
    P(\eta_{D_{\mathrm{Rx}}}|r) &\approx 
    \begin{cases}
        \dfrac{e^{-\frac{(\ln \eta_{D_{\mathrm{Rx}}} + \mu)^2}{2\sigma^2}}}{\mathcal{F}(1)\sqrt{2\pi}\eta_{D_{\mathrm{Rx}}}\sigma}, & \text{for } \eta_{D_{\mathrm{Rx}}} \in [0,1], \\
        0, & \text{otherwise},
    \end{cases}
    \label{eq:p_strong_definition}
\end{align}
where $\mathcal{F}(1)$ is the cumulative distribution function (CDF) at $\eta_{D_{\mathrm{Rx}}}=1$, 
$\mu$ and $\sigma$ are two parameters of the log-normal distribution, and can be computed as a function of beam displacement $r$ as in \cite[(B20, B22)]{karakosta2025free}.

\subsubsection{Fiber-coupling Efficiency}
\label{sec:fiber_coupling}
The fiber-coupling efficiency is the fraction of optical signals from the receiver telescope that are focused and matched into a single-mode fiber. The coupling-efficiency PDF is given as \cite[(12)]{karakosta2025free}
\begin{align}
    f_{\eta_{\mathrm{SMF}}} \left( \eta_{\mathrm{SMF}} \right) = \frac{1}{\eta_{\mathrm{SMF}}}\,
    f_\xi\!\left[\ln\!\left(\frac{\eta_0\,\eta_\chi }{\eta_{\mathrm{SMF}}}\right)\right],
    \label{eq:smf_pdf}
\end{align}
where $\eta_0$ is the maximum coupling efficiency in the absence of turbulence, which is calculated as a function of the obstruction ratio 
$\alpha_{\mathrm{obs}}$ as in \cite[(C11)]{karakosta2025free}. Note that the obstruction in this case is caused by the telescope's support structures, such as a secondary mirror. Moreover, $\eta_\chi$ denotes the efficiency due to atmospheric turbulence, which is given by \cite[(C13)]{karakosta2025free}
\begin{align}
    \eta_\chi \approx \exp\left[ -\sigma_\chi^2 \right],
\end{align}
where $\sigma_\chi^2$ is the log-amplitude variance, and is given in \cite[(C14)]{karakosta2025free}. Finally, $f_\xi$ is the PDF of the wavefront residual error after adaptive optics (AO), which is given as \cite[(11)]{karakosta2025free}
\begin{align}
    f_\xi(\xi) &= \frac{1}{\pi}\int_0^{\infty}
    \frac{\cos\!\left[\sum_j\frac{1}{2}\arctan\!\left(2\gamma_j^2\langle b_j^2\rangle u\right)-\xi u\right]}{\prod_j\left[1+4u^2\gamma_j^4\langle b_j^2\rangle^2\right]^{1/4}} \mathrm{d} u,
    \label{eq:residual_phase_pdf}
\end{align}
where $b_j$ is the aberration coefficient for the $j$-th phase aberration and can be calculated as in \cite[(C24)]{karakosta2025free}. Moreover, $\gamma_j$ is the attenuation factor of AO for the $j$-th phase aberration. Given the order of AO correction $N_\text{AO}$, $\gamma_j = 1$ if $j > N_\text{AO}$; otherwise, its calculation is given in \cite[(C35)]{karakosta2025free}.

\section{Qiskit Simulation \& Analysis}
\label{sec:simul_analysis}

\subsection{Qiskit-based Simulation}
\label{sec:simulation}
\begin{figure}[t]
  \centering
    \includegraphics[width=0.9\columnwidth]{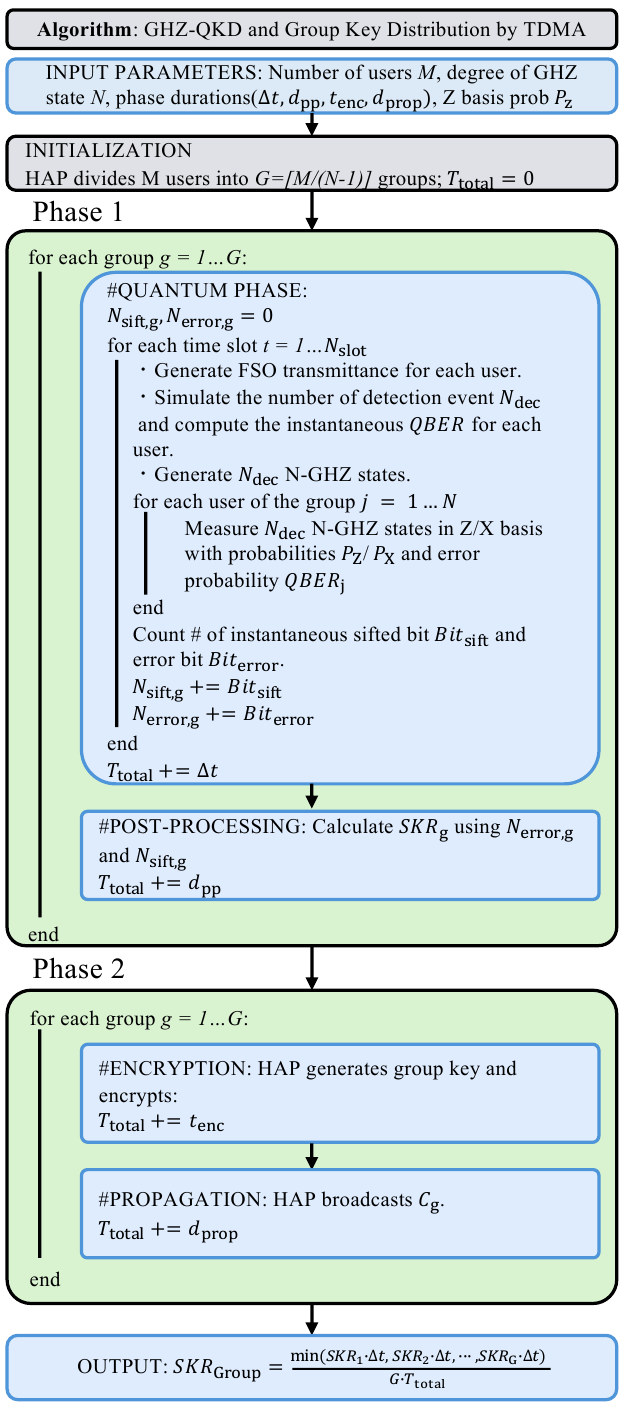}
    \caption{Algorithm for the proposed $N$-GHZ-QKD and group key distribution.}
    \label{fig:algorithm}
\end{figure}
To evaluate the feasibility and performance of the proposed $N$-GHZ QKD system, we develop a comprehensive simulation framework that integrates an FSO channel model with a Qiskit-based implementation of the $N$-GHZ protocol as depicted in Fig.~\ref{fig:algorithm}.
For each simulation run, the instantaneous transmittance values for each user in the group are generated using the FSO channel model. Subsequently, these values are used to calculate the number of photon-detection events ($N_\text{dec}$) and the corresponding instantaneous quantum bit error rate ($\text{QBER}_{j}$) for each user. These parameters then serve as direct inputs to the Qiskit-based simulation. To emulate a realistic quantum noise environment in Qiskit, a Pauli error model is applied to the qubits of the generated $N$-GHZ states based on the computed QBER.
Finally, each user performs photon measurements according to the basis selection probabilities, $P_\text{Z}$ and $P_X$. The framework computes the number of sifted bits ($N_\text{sift}$) and erroneous bits ($N_\text{error}$), which are used to evaluate the achievable group secret key rate.

\subsection{Theoretical Secret Key Rate (SKR)}
\label{sec:analysis}

The group SKR, which is defined as the number of group secret bits shared per second, is expressed as
\begin{align}
    \label{eq:group_SKR}
    \mathrm{SKR}_\text{G} \left( P_\text{Z}, M, N \right)
    &= \frac{\min\limits_{g \in \{1, \dots, G\}} \left[ \mathrm{AKR}_g \left( P_\text{Z}, N \right)  \Delta t \right]}{G  \left( \Delta t + d_{\mathrm{pp}} + t_{\mathrm{enc}} + d_{\mathrm{prop}} \right)},
\end{align}
where $G = \left\lceil \frac{M}{N-1} \right\rceil$ is the number of subgroups, $\Delta t$ is the quantum phase delay, $d_{\mathrm{pp}}$ is the post-processing delay, $t_{\mathrm{enc}}$ is the encryption time, and $d_{\mathrm{prop}}$ is the propagation delay.
The asymptotic key rate (AKR) can be calculated as \cite{murta2020quantum}
\begin{align}
    \label{eq:AKR}
    \mathrm{AKR}_g \left( P_\text{Z}, N \right) = \mathcal{R}  \langle Y_N^{\left( g \right)} \rangle  r_{\infty}^{(g)}  P_{\mathrm{sift}}(P_\text{Z}, N),
\end{align}
where $\mathcal{R}$ is the pulse repetition rate, $\langle Y_N^{\left( g \right)} \rangle$ is the probability that clicks occur at all users' detectors, given as
\begin{align}
    \label{eq:prob_click_events}
    \langle Y_N^{\left( g \right)} \rangle \!\! &= \!\!\! \int_{0}^{1} \!\!\!\! \dots \!\! \int_{0}^{1} \!\! \displaystyle\prod_{i=1}^{N-1} \!\! Y \!\! \left( \eta_i \right) \!\! f_{\eta_{1}} \!\! \left( \eta_{1} \right) \! \dots \! f_{\eta_{{N-1}}} \!\! \left( \eta_{{N-1}} \right) \mathrm{d}\eta_{{1}} \! \dots \! \mathrm{d}\eta_{{N-1}},
\end{align}
where $Y \left( \eta \right) = 1 - \left( 1 - p_\text{dark} \right) \left(1 - \eta \right)$ is the detection probability given the transmittance $\eta$, and $p_\text{dark}$ is the dark count rate. As each $\eta_i$ is independent of the others and $ Y \left( \eta_i \right)$ is linear in $\eta_i$, we can rewrite \eqref{eq:prob_click_events} as
\begin{align}
    \label{eq:prob_click_events_2}
    \langle Y_N^{\left( g \right)} \rangle = \displaystyle\prod_{i=1}^{N-1} Y \left( \langle \eta_i \rangle \right), 
\end{align}
where $\langle \eta_i \rangle = \int_{0}^{1} \eta_i f_{\eta_i} \left( \eta_i \right) \mathrm{d}\eta_i$ is the effective channel transmittance, which is solved numerically in our framework. Moreover, $r_{\infty}$ is the fraction of secret bits and the number of shared states, which is given as \cite{pickston2023conference}
\begin{align}
    \label{eq:r_infty}
    r_{\infty}^{(g)} = 1 - h \left( Q_{\mathrm{x}} \right) - \max_{ 1 \leq i \leq N-1 } \!\! h \left( \langle Q_{\mathrm{AB}_i} \rangle \right),
\end{align}
where $h \left( \cdot \right)$ is the binary entropy function, $Q_{\mathrm{x}}$ is the phase error, and $\langle Q_{\mathrm{AB}_i} \rangle$ is the average quantum bit-error rate between Alice and $i$-th Bob. Therein, we assume that the error rate is caused by two main factors, i.e., dark count and intrinsic detector errors. Hence, $\langle Q_{\mathrm{AB}_i} \rangle$ can be expressed as \cite{kaminaga2026qiskit}
\begin{align}
    \label{eq:Q_ABi}
    \langle Q_{\mathrm{AB}_i} \rangle \!\! &= \!\! 
    \frac{
        \displaystyle\int_{0}^{1} \!\!\!\! \dots \!\! \displaystyle\int_{0}^{1} \!\! E \left( \overrightarrow{\bm{\eta}} \right) \!\! f_{\eta_{1}} \!\! \left( \eta_{1} \right) \! \dots \! f_{\eta_{{N-1}}} \!\! \left( \eta_{{N-1}} \right) \mathrm{d}\eta_{{1}} \! \dots \! \mathrm{d}\eta_{{N-1}} 
    }{
        \langle Y_N^{\left( g \right)} \rangle
    } \nonumber \\
    &= \frac{ E \left( \langle \overrightarrow{\bm{\eta}} \rangle \right)}{\langle Y_N^{\left( g \right)} \rangle},
\end{align}
where $\overrightarrow{\bm{\eta}} = \left[ \eta_1, \dots, \eta_{N-1} \right]$ is the vector presenting the transmittance of each Bob, 
$\langle \overrightarrow{\bm{\eta}} \rangle = \left[ \langle \eta_1 \rangle, \dots, \langle \eta_{N-1} \rangle \right]$, $E \left( \overrightarrow{\bm{\eta}} \right)$ is the erroneous detection probability per pulse conditional on $\overrightarrow{\bm{\eta}}$, and can be computed as
\begin{align}
    \label{eq:err_dec_prob}
    E \left( \overrightarrow{\bm{\eta}} \right) = e_0 \left( Y_N - Y_{\mathrm{A} \mathrm{B}_i} \right) + e_\mathrm{dec} Y_{\mathrm{A} \mathrm{B}_i},
\end{align}
Moreover, $Y_N = \prod_{i=1}^{N-1} Y \left( \eta_i \right)$, $Y_{\mathrm{A} \mathrm{B}_i} = \eta_i \frac{Y_N}{ Y \left( \eta_i \right)}$, $e_0 = 0.5$ is the error rate due to dark counts, $e_\mathrm{dec}$ is the intrinsic detector error rate. In addition, $P_{\mathrm{sift}}$ is the sifting probability, which is expressed as 
\begin{align}
    \label{eq:sift_prob}
    P_{\mathrm{sift}}
    = 
    \begin{cases} 
    P_\text{Z}^2 + (1-P_\text{Z})^2 & \text{if } N = 2, \\
    P_\text{Z}^N & \text{if } N \ge 3,
    \end{cases}
\end{align}
where $P_\text{Z}$ denotes the probability of selecting the Z basis.

\section{Results and Discussions}
\label{sec:results}
This section presents and discusses the performance of the proposed HAP-assisted $N$-GHZ-QKD system.
Unless otherwise stated, the parameters are as follows.
At the HAP (Alice), the HAP altitude $H = 20$ km,
beam waist $W_0 = 0.10$ m, optical wavelength $\lambda = 1550$ nm, 
and pulse repetition rate $\mathcal{R} = 10^9$ pulses/s.
At the ground users (Bob),
number of users $M = 6$,
zenith angle $\theta_z = 0^\circ$, 
receiver aperture $D_{\mathrm{Rx}} = 0.9$ m, 
telescope obstruction ratio $\alpha_{\mathrm{obs}} = 0.1$, 
ground user altitude $h_0 = 20$ m, 
order of AO correction $N_\text{AO} = 10$,
intrinsic detector errors $e_{\mathrm{dec}} = 3.3\%$,
background detection probability $p_{\mathrm{dark}} = 10^{-4}$,
and phase error $Q_{\mathrm{x}} = 2\%$.
\textit{Regarding the FSO channel,} 
rms wind speed $v_\text{wind} = 10$ m/s,
ground-level refractive index $C_{\mathrm{n}}^2 \left( 0 \right) = 10^{-15}$ m$^{-2/3}$,
atmospheric transmittance $\eta_{\mathrm{atm}} = 0.9$,
angular pointing error $\theta_{\rm pe} = 1~\mu$rad, 
and tracking efficiency $\eta_{\mathrm{tr}} = 1$.
Other parameters can be found in \cite{karakosta2025free}.

\begin{figure}[t]
  \centering
    \includegraphics[width=\columnwidth]{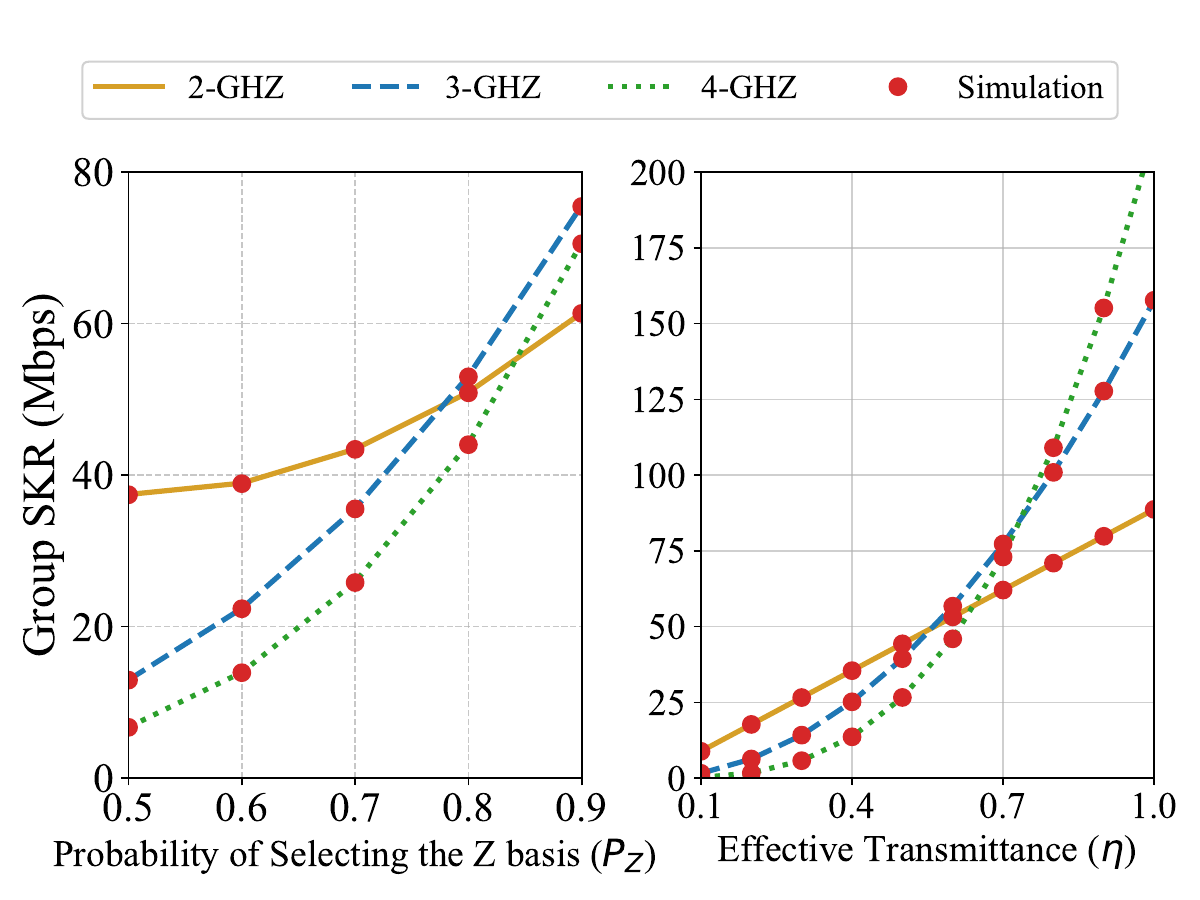}
    \caption{Group SKR versus the probability of selecting Z basis $P_\text{Z}$ (left), and effective transmittance (right).}
    \label{fig:result1_2}
\end{figure}

First, we investigate the performance of different $N$-GHZ protocols as a function of the Z-basis selection probability, $P_\text{Z}$, as shown in Fig.~\ref{fig:result1_2}. As observed, the theoretical results closely follow the simulation results, confirming the correctness of our analytical framework. Moreover, the high-order $N$-GHZ protocols outperform the 2-GHZ protocol as $P_\text{Z}$ increases. This is because the sifting probability scales as cubic/quartic in $P_\text{Z}$ when $N=3/4$. Therefore, selecting a relatively large value of $P_\text{Z}$ (around $0.8$ or $0.9$) is advantageous for $3/4$-GHZ protocols. The associated trade-off is that more time is required to accumulate a sufficient secondary basis (X) for error estimation. However, this limitation can be alleviated by increasing the duration of the quantum communication phase \cite{lo2005efficient}.

Another important factor that can affect the SKR of $N$-GHZ protocols is the effective channel transmittance, as illustrated in Fig.~\ref{fig:result1_2}. It is observed that as the effective transmittance increases, the performance of high-order $N$-GHZ protocols surpasses that of lower-order ones. This is due to the increase in the probability that clicks occur at all users’ detectors. From this result, we can select the GHZ order to maximize the group SKR. For example, when the effective transmittance is $0.6$, we can select the $3$-GHZ protocol.

\begin{figure}[t]
  \centering
    \includegraphics[width=0.9\columnwidth]{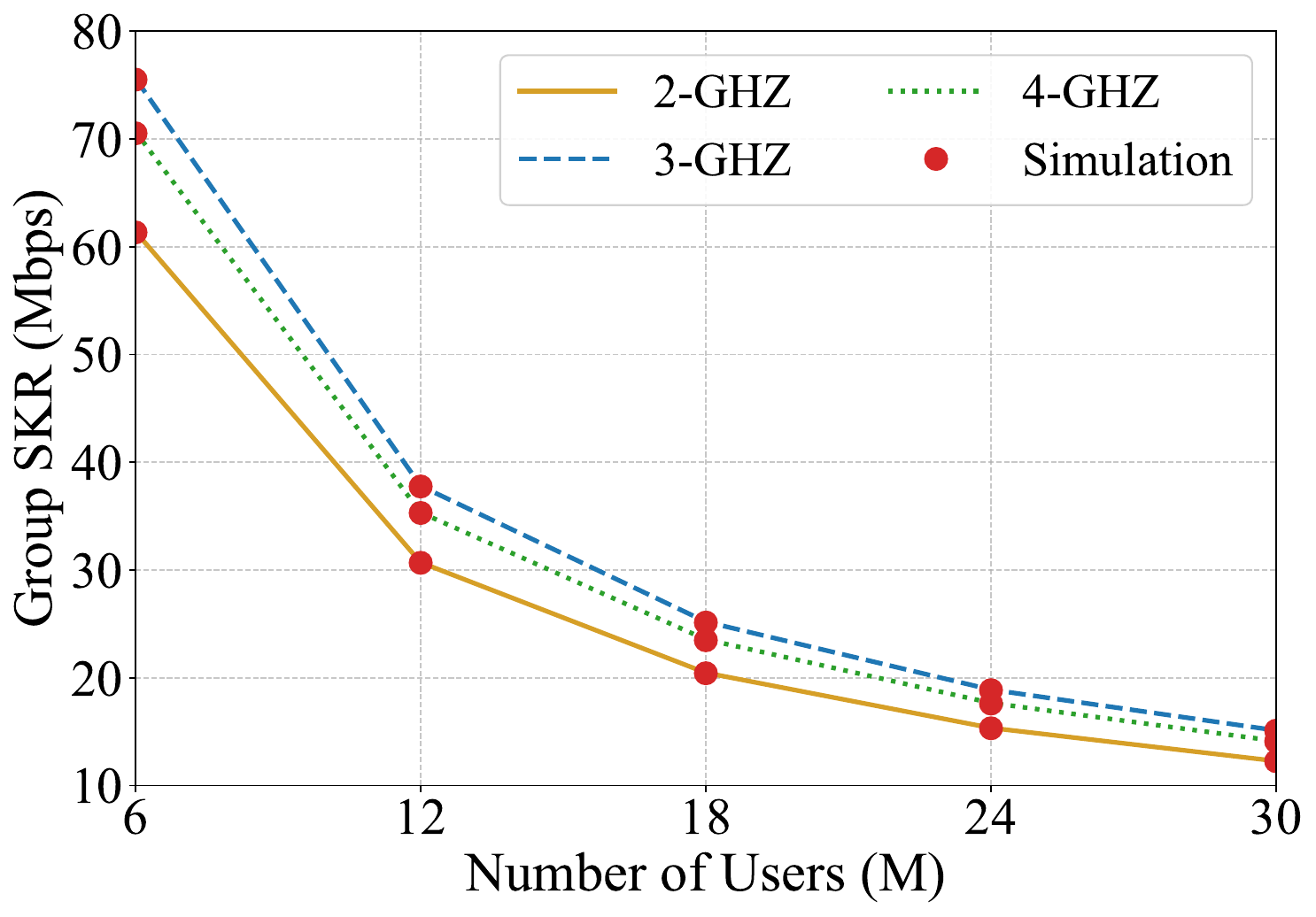}
    \caption{Group SKR versus the total number of users ($M$)}
    \label{fig:result3}
\end{figure}
\begin{figure}[t]
  \centering
    \includegraphics[width=0.9\columnwidth]{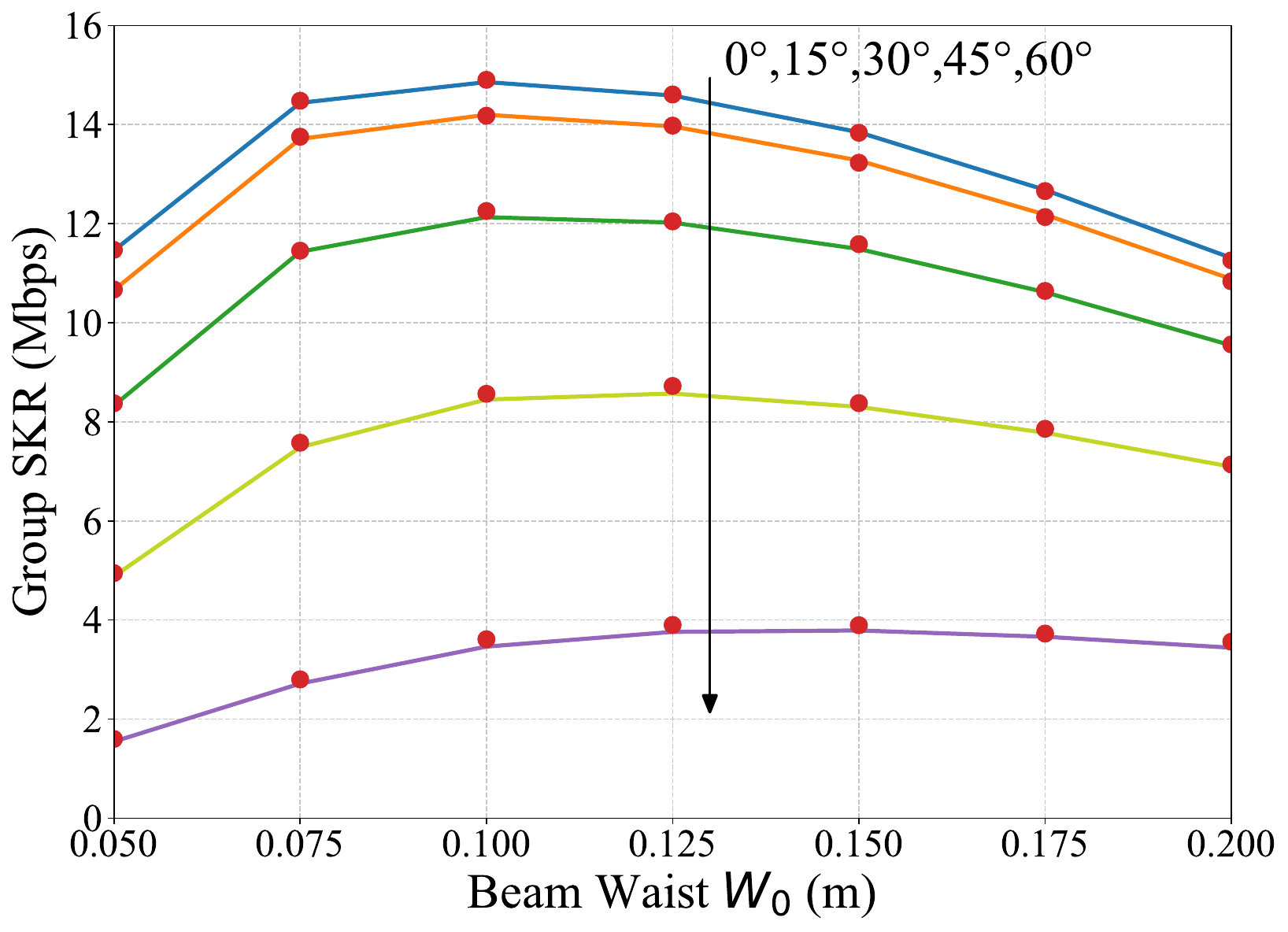}
    \caption{Group SKR versus the $W_0$ for various zenith angles using the 3-GHZ}
    \label{fig:result4}
\end{figure}
Next, we evaluate the system performance when the number of users increases, as shown in Fig.~\ref{fig:result3}. The results indicate that the group secret key rate (SKR) decreases with the number of users for all considered protocols. This happens because a higher $M$ forces the HAP to divide users into more groups ($G$), thereby increasing the total TDMA cycle duration. From this result, we can determine the maximum number of users for a given group SKR level. For instance, if the required group SKR level is $20$~Mbps, the maximum number of users is $18$.

Finally, Fig.~\ref{fig:result4} illustrates the performance of the 3-GHz-QKD protocol versus the transmission beam waist. The parameters are adopted from \cite{karakosta2025free}. Moreover, we assume that all users are located in the same area and that their zenith angles are nearly identical. As shown, there is an optimal beam waist that maximizes the group SKR. This is because a small beam waist is more susceptible to pointing error, while a big one suffers from geometric loss. From this result, we can select the optimal beam waist for users' zenith angles. For instance, the optimal beam waist should be set to $0.1$, $0.125$, and $0.15$, when the zenith angles are $0$, $45$, and $60$ degrees, respectively.

\section{Conclusions}
\label{sec:conclusions}
This paper proposed and evaluated a GHZ-based QKD framework for HAP-assisted 6G wireless networks. To assess the feasibility and practical performance of the proposed architecture, we developed both a comprehensive Qiskit-based simulation platform and an analytical SKR framework that incorporates the major impairments encountered in FSO channels, including atmospheric attenuation, turbulence-induced fading, and pointing errors. The strong agreement observed between simulation and theoretical results validates the accuracy and reliability of the proposed modeling framework. Using the developed framework, we systematically analyzed the impact of several critical system parameters on network performance, including the GHZ entanglement order, the effective channel transmittance, and the number of users. The results demonstrate trade-offs among scalability, transmission reliability, and secret-key generation efficiency in HAP-assisted multipartite quantum communication systems.

\bibliographystyle{IEEEtran}
\bibliography{references}

@article{karakosta2025free,
  author   = {Karakosta-Amarantidou, Ilektra and Yehia, Raja and Schiavon, Matteo},
  title    = {Free-space model for a balloon-based quantum network},
  journal  = {Phys. Rev. Res.},
  volume   = {7},
  number   = {2},
  pages    = {023199},
  year     = {May 2025}
}

@article{epping2017multi,
  title={Multi-partite entanglement can speed up quantum key distribution in networks},
  author={Epping, Michael and Kampermann, Hermann and Macchiavello, Chiara and Bru{\ss}, Dagmar},
  journal={New J. Phys.},
  volume={19},
  number={9},
  pages={093012},
  year={Sep. 2017}
}

@article{pickston2023conference,
  title={Conference key agreement in a quantum network},
  author={Pickston, Alexander and others},
  journal={npj Quantum Inf.},
  volume={9},
  number={1},
  pages={82},
  year={Aug. 2023}
}

@article{murta2020quantum,
  title={Quantum conference key agreement: A review},
  author={Murta, Gl{\'a}ucia and Grasselli, Federico and Kampermann, Hermann and Bru{\ss}, Dagmar},
  journal={Adv. Quantum Technol.},
  volume={3},
  number={11},
  pages={2000025},
  year={Nov. 2020},
  publisher={Wiley Online Library}
}

@article{hegland2006survey,
  title={A survey of key management in ad hoc networks},
  author={Hegland, Anne Marie and Winjum, Eli and Mjolsnes, Stig F and Rong, Chunming and Kure, Oivind and Spilling, Pal},
  journal={IEEE Communications Surveys \& Tutorials},
  volume={8},
  number={3},
  pages={48--66},
  year={2006},
  publisher={IEEE}
}

@inproceedings{kaminaga2026qiskit,
  title={Qiskit-Based Simulation of {HAP}-Enabled Multi-User Entanglement {QKD} over {FSO} Channels},
  author={Kaminaga, Yuma and Le, Hoang D and Nguyen, Cuong T and Pham, Anh T},
  booktitle={Proc. Int. Conf. Artif. Intell. Inf. Commun. (ICAIIC)},
  pages={059--063},
  year={2026}
}

@article{chu2021feasibility,
  title={Feasibility of quantum key distribution from high altitude platforms},
  author={Chu, Yi and Donaldson, Ross and Kumar, Rupesh and Grace, David},
  journal={Quantum Science \& Technology},
  volume={6},
  number={3},
  pages={035009},
  year={2021},
  publisher={IOP Publishing}
}

@article{vu2020design,
  title={Design and performance of relay-assisted satellite free-space optical quantum key distribution systems},
  author={Vu, Minh Quang and Pham, Thanh V and Dang, Ngoc T and Pham, Anh T},
  journal={IEEE Access},
  volume={8},
  pages={122498--122510},
  year={2020},
  publisher={IEEE}
}

@inproceedings{alshaer2024exploring,
  title={Exploring quantum key distribution for secure communication in high-altitude platforms},
  author={Alshaer, Nancy and Ismail, Tawfik},
  booktitle={2024 24th International Conference on Transparent Optical Networks (ICTON)},
  pages={1--5},
  year={2024},
  organization={IEEE}
}

@article{trinh2025optical,
  title={Optical RISs improve the secret key rate of free-space QKD in HAP-to-UAV scenarios},
  author={Trinh, Phuc V and Sugiura, Shinya and Xu, Chao and Hanzo, Lajos},
  journal={IEEE Journal on Selected Areas in Communications},
  year={2025},
  publisher={IEEE}
}

@article{lo2005efficient,
  title={Efficient quantum key distribution scheme and a proof of its unconditional security},
  author={Lo, Hoi-Kwong and Chau, Hoi Fung and Ardehali, Mohammed},
  journal={J. Cryptol.},
  volume={18},
  number={2},
  pages={133--165},
  year={Apr. 2005},
  publisher={Springer}
}

@article{scarani2009security,
  title={The security of practical quantum key distribution},
  author={Scarani, Valerio and Bechmann-Pasquinucci, Helle and Cerf, Nicolas J and Du{\v{s}}ek, Miloslav and L{\"u}tkenhaus, Norbert and Peev, Momtchil},
  journal={Reviews of modern physics},
  volume={81},
  number={3},
  pages={1301--1350},
  year={2009},
  publisher={APS}
}

\end{document}